\documentclass[aip,rsi,amsmath,amssymb,reprint,superscriptaddress]{revtex4-1}
\usepackage{graphicx}
\usepackage{dcolumn}
\usepackage{bm}
\begin{document}
\title{Polarization- and frequency-tunable microwave circuit for selective excitation of nitrogen-vacancy spins in diamond}
\author{Johannes Herrmann}
\affiliation{School of Fundamental Science and Technology, Keio University, 3-14-1 Hiyoshi, Kohoku-ku, Yokohama 223-8522, Japan}
\author{Marc A. Appleton}
\affiliation{School of Fundamental Science and Technology, Keio University, 3-14-1 Hiyoshi, Kohoku-ku, Yokohama 223-8522, Japan}
\author{Kento Sasaki}
\affiliation{School of Fundamental Science and Technology, Keio University, 3-14-1 Hiyoshi, Kohoku-ku, Yokohama 223-8522, Japan}
\author{Yasuaki Monnai}
\affiliation{School of Fundamental Science and Technology, Keio University, 3-14-1 Hiyoshi, Kohoku-ku, Yokohama 223-8522, Japan}
\author{Tokuyuki Teraji}
\affiliation{National Institute for Materials Science, 1-1 Namiki, Tsukuba, Ibaraki 305-0044, Japan}
\author{Kohei M. Itoh}
\email{kitoh@appi.keio.ac.jp}
\affiliation{School of Fundamental Science and Technology, Keio University, 3-14-1 Hiyoshi, Kohoku-ku, Yokohama 223-8522, Japan}
\affiliation{Spintronics Research Center, Keio University, 3-14-1 Hiyoshi, Kohoku-ku, Yokohama 223-8522, Japan}
\author{Eisuke Abe}
\email{e-abe@keio.jp}
\affiliation{Spintronics Research Center, Keio University, 3-14-1 Hiyoshi, Kohoku-ku, Yokohama 223-8522, Japan}
\date{\today}

\begin{abstract}
We report on a planar microwave resonator providing arbitrarily polarized oscillating magnetic fields that enable selective excitation of the electronic spins of nitrogen-vacancy (NV) centers in diamond.
The polarization plane is parallel to the surface of diamond, which makes the resonator fully compatible with (111)-oriented diamond.
The field distribution is spatially uniform in a circular area with a diameter of 4~mm, and a near-perfect circular polarization is achieved.
We also demonstrate that the original resonance frequency of 2.8~GHz can be varied in the range of 2$-$3.2~GHz by introducing varactor diodes that serve as variable capacitors.
\end{abstract}
\maketitle

\begin{figure}
\begin{center}
\includegraphics{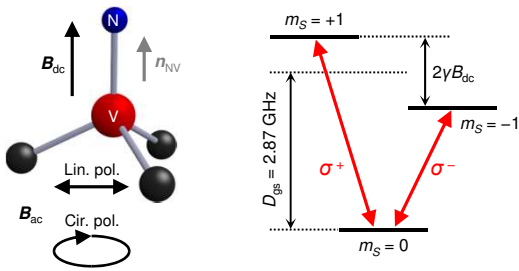}
\caption{(left) Schematic of the NV center.
The center red sphere represents a vacancy, the top blue sphere a nitrogen atom, and the three black spheres carbon atoms.
(right) Energy levels of the negatively charged NV center with the static magnetic field $\bm{B}_{\mathrm{dc}}$ applied along the quantization axis $\bm{n}_{\mathrm{NV}}$.
}
\label{fig1}
\end{center}
\end{figure}

In recent years, the nitrogen-vacancy (NV) center in diamond has emerged as a promising platform for quantum information processing and nanoscale metrology.~\cite{CH13,TBB+13,HGP+13,HSH13,SCLD14,RTH+14,IW14}
At the heart of both technologies lies an exquisite control of the NV electronic spin.~\cite{DFF+13}
The ground state of the negatively charged NV center is an $S$ = 1 spin triplet with the $m_S$ = $\pm$1 sublevels
lying $D_{\mathrm{gs}}$ = 2.87~GHz above the $m_S$ = 0 sublevel under zero magnetic field.
The external static magnetic field $\bm{B}_{\mathrm{dc}}$ applied parallel to the quantization axis $\bm{n}_{\mathrm{NV}}$ (along one of the $\langle$111$\rangle$ crystallographic axes)
further splits the $m_S$ = $\pm$1 levels by 2$\gamma B_{\mathrm{dc}}$,
with $\gamma$ = 28~GHz/T being the gyromagnetic ratio of the NV electronic spin [see Fig.~\ref{fig1}].

Initialization, control, and readout of this three-level V system are accomplished by an optically detected magnetic resonance (ODMR) technique,
in which optical pumping by a green ($\sim$500~nm) laser serves to initialize and read out the NV spin
and short pulses of oscillating magnetic fields $\bm{B}_{\mathrm{ac}}$ ($\sim$2.87~GHz) drive it into an arbitrary quantum-mechanical superposition.
In ideal magnetic resonance experiments, $\bm{B}_{\mathrm{ac}}$ and $\bm{B}_{\mathrm{dc}}$ ($\parallel \bm{n}_{\mathrm{NV}}$) should be orthogonal,~\cite{S90,AJ01}
whereas in reality the direction of $\bm{B}_{\mathrm{ac}}$ at the location of the target NV center is hard to control or even know about.
This is because a metal wire or a microfabricated stripline, commonly used for experiments with NV centers, generates $\bm{B}_{\mathrm{ac}}$ that is highly dependent on the positions.
Moreover, $\bm{B}_{\mathrm{ac}}$ from these sources is linearly polarized, while the NV spin has clear transition selection rules;
The $m_S$ = 0 $\leftrightarrow$ 1 ($-$1) transition is driven by $\sigma^{+}$ ($\sigma^{-}$) circularly polarized fields.
To fully exploit the $S$ = 1 nature of the NV spin for quantum information and metrology applications,~\cite{ZWH+14,ALHB14,SKM+16,ANW+15}
it is highly desired to have a reliable means to generate arbitrarily polarized microwave fields.

A few configurations for polarization-controlled $\bm{B}_{\mathrm{ac}}$ have been adopted for the NV system.
A popular one is a pair of crisscrossed striplines, which generates desired fields only beneath the crossing point.~\cite{ASMB07,LBN+14}
A pair of parallel striplines may be a simpler alternative.~\cite{MMRG15}
In both examples, however, it is by design difficult or impossible to make the polarization plane perpendicular to $\bm{n}_{\mathrm{NV}}$ and thus to $\bm{B}_{\mathrm{dc}}$.

In this paper, we present a simple planar resonator circuit providing spatially uniform, arbitrarily polarized, and {\it in-plane} microwave magnetic fields.
With a (111)-oriented diamond and $\bm{B}_{\mathrm{dc}}$ applied perpendicular to the diamond surface,
both $\bm{B}_{\mathrm{dc}}$ $\perp$ $\bm{B}_{\mathrm{ac}}$ and $\bm{n}_{\mathrm{NV}}$ $\perp$ $\bm{B}_{\mathrm{ac}}$ are readily realized.
This results in near-perfect selective excitation of the NV spin. 
In (111)-oriented synthetic diamond, it has been reported that the NV centers can be preferentially oriented along the [111] crystallographic axis.~\cite{EHC+12,MTZ+14,LTT+14,FDM+14}
Attracting enormous attention, such a special diamond substrate will lead to enhanced sensitivity in metrology as well as atomically precise quantum information devices.
Our resonator circuit design will be fully compatible with these applications.

\begin{figure}
\begin{center}
\includegraphics{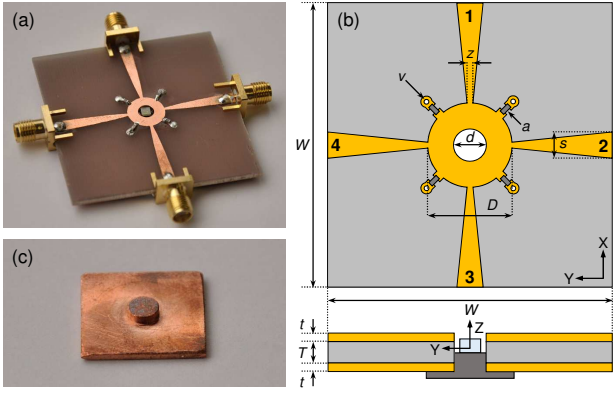}
\caption{(a) Photo of the fabricated copper stripline with capacitors and diamond sample mounted.
(b) Design parameters and port numbers.
$D$ = 10~mm (outer resonator diameter),
$d$ = 4~mm (diameter of the hole to place a diamond sample),
$s$ = 3.4~mm (width of stripline at the SMA connectors),
$z$ = 0.5~mm (width of stripline at the end of the impedance transition),
$v$ = 1.4~mm (diameter of via holes connecting the top and back ground planes),
$T$ = 1.6~mm (thickness of epoxy glass),
$t$ = 35~$\mu$m (thickness of copper films).
$a$ represents a standard 0603-footprint for a high frequency capacitor with the capacitance $C_0$ = 1.1~pF (SN: ATC600S).
To attain a 50-$\Omega$ impedance matching, the line width $s$ is transformed to the value of $z$ with an exponential curve; shown simplified.
(c) Photo of the copper socket, serving as a stand for the diamond sample.
}
\label{fig2}
\end{center}
\end{figure}

Figure~\ref{fig2}(a) is a photograph of the fabricated microwave circuit consisting of a ring cavity connected to four striplines.
The cavity is also loaded with four 1.1-pF capacitors.
To establish electrical contacts between the capacitors and the ground plane on the back side, four via holes are manufactured.
The circular structure and the four symmetrically distributed capacitors form a microwave resonator, allowing a high magnetic field amplitude.
Only Ports 1 and 2 are used to feed in the microwave power, and the other two ports are connected to 50-$\Omega$ terminators.
Nonetheless, all the four striplines are required to preserve the symmetry of the structure, which turns out to be essential to achieve a high degree of microwave polarization.
The design parameters are summarized in Fig.~\ref{fig2}(b).

\begin{figure}
\begin{center}
\includegraphics{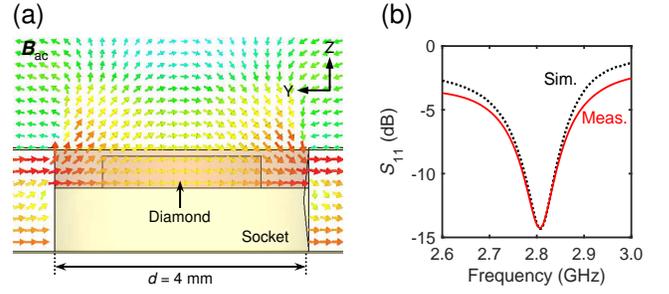}
\caption{(a) Simulated distribution of $\bm{B}_{\mathrm{ac}}$ in the YZ plane at 2.87~GHz and at its initial phase.
The arrows indicate the directions of $\bm{B}_{\mathrm{ac}}$, and their color reflects the relative field strength (strong in red and weak in cyan).
(b) Measured (solid line) and simulated (dotted line) $S_{11}$ parameters.}
\label{fig3}
\end{center}
\end{figure}

To verify and optimize the design, we used a three-dimensional electromagnetic field simulation software CST MICROWAVE STUDIO\textsuperscript{\textregistered}.
The dimensions of a simulated diamond (2.5 $\times$ 2.5~$\times$ 0.5~mm$^3$) are set to the size of a diamond sample used in ODMR experiments to be presented below.
A cross-sectional distribution of simulated $\bm{B}_{\mathrm{ac}}$ is shown in Fig.~\ref{fig3}(a).
The sample is mounted on a copper socket [Fig.~\ref{fig2}(c)], which is electrically connected to the ground with its height adjusted according to the sample size.
In this way, $\bm{B}_{\mathrm{ac}}$ inside of the sample can be made uniform and parallel to the sample surface.
Figure~\ref{fig3}(b) is a comparison of simulated and measured reflection coefficients $S_{11}$.
We confirm a good agreement between them, and obtain the resonance frequency $f_{\mathrm{res}}$ of 2.806~GHz and the full-width at half maximum of 165~MHz (quality factor $Q$ of 17).
We note that the introduction of an objective lens, indispensable for ODMR spectroscopy, shifts $f_{\mathrm{res}}$ toward 2.87~GHz due the proximity of its metal housing.
On the other hand, $Q$ is not strongly affected by the objective lens, and the resonator secures a bandwidth sufficient for low-field ODMR.

In analyzing the simulation, we find that the mounted capacitors dominate the total capacitance.
Combined with resistance and inductance arising from the structure, $f_{\mathrm{res}}$ and $Q$ are determined.
As will be shown later, varactor diodes can be used as variable capacitors, and this provides a convenient means to tune $f_{\mathrm{res}}$ between 2 and 3.2~GHz, far beyond the range allowed by the $Q$ factor.

\begin{figure}
\begin{center}
\includegraphics{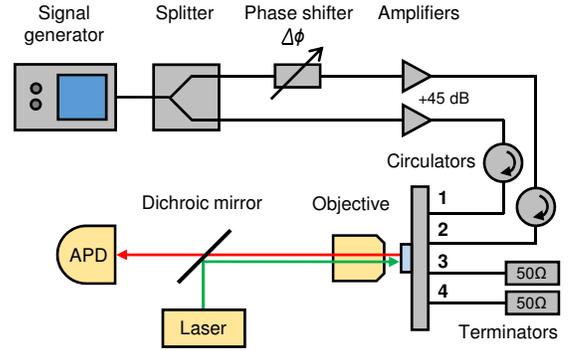}
\caption{Schematic of the experimental setup consisting of the microwave (gray) and optics (yellow) parts.
APD: avalanche photodiode.
}
\label{fig4}
\end{center}
\end{figure}

We now test the performance of our resonator through ODMR on the NV spins.
Our experimental setup is schematically shown in Fig.~\ref{fig4}.
A 515-nm green laser excites the NV electrons and the red photons emitted from the NV center are counted with an avalanche photodiode.
To create two phase-shifted microwave signals, a programmable phase shifter is inserted in one of the electrical lines after a power splitter.
The phase offset $\Delta \phi$ is variable from $-$180$^{\circ}$ to 180$^{\circ}$ with a resolution of 6~bits.
The effect of the group delay $\tau_g$, arising from the difference in the electrical lengths of the two paths, is carefully eliminated.~\cite{delay}
Circulators protect the setup from reflected microwave power.
As mentioned above, Ports 3 and 4 are terminated.
A type Ib (111) diamond substrate, 2.5 $\times$ 2.5 $\times$ 0.5 mm$^3$ in size, is mounted at the center of the resonator,
and a permanent magnet placed at the back of the copper socket provides $B_{\mathrm{dc}}$ = 1.9~mT parallel to the [111] crystallographic axis, which is the quantization axis of interest.

\begin{figure}
\begin{center}
\includegraphics{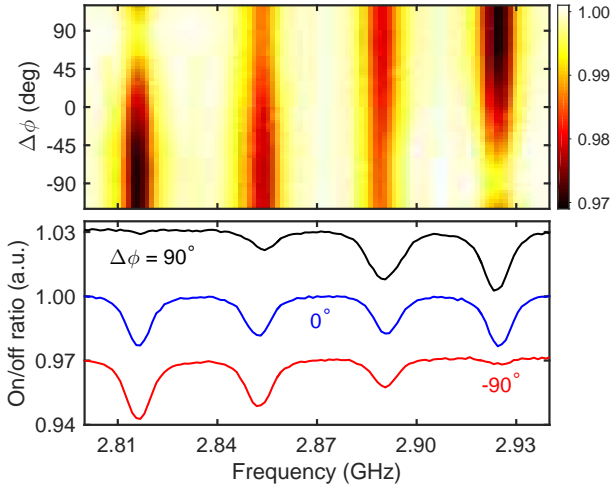}
\caption{Demonstration of polarization tunability.
(top) CW ODMR spectra as a function of $\Delta \phi$ ($-$120$^{\circ}$ $\leq \Delta \phi \leq$ 120$^{\circ}$) at $B_{\mathrm{dc}}$ = 1.9~mT.
The on/off ratio is color coded.
(bottom) CW ODMR spectra at $\Delta \phi$ = 90$^{\circ}$ (black, shifted by +0.03), 0$^{\circ}$ (blue) and $-$90$^{\circ}$ (red, shifted by $-$0.03).
}
\label{fig5}
\end{center}
\end{figure}

Figure~\ref{fig5} shows continuous wave (CW) ODMR spectra of ensemble NV centers as a function of $\Delta \phi$.
As the microwave frequency is swept, the fluorescence decreases at the resonances, creating four dips in the optical contrast (shown as a ratio of the photon counts when the microwave is on and off).
The outer two dips are from the quantization axis aligned with $\bm{B}_{\mathrm{dc}}$ $\parallel$ [111],
whereas the inner two from three degenerate, non-aligned quantization axes experiencing a weaker static magnetic field of $B_{\mathrm{dc}} \cos(109.5^{\circ})$.
Here, we focus on the outer two dips, as they allow straightforward analysis.
It is clearly observed that the $m_S$ = $\pm$1 transitions can be selectively or simultaneously excited by appropriately choosing $\Delta \phi$;
At $\Delta \phi$ = 90$^{\circ}$ ($-$90$^{\circ}$), only the $m_S$ = 0 $\leftrightarrow$ 1 ($-$1) transition is excited, corresponding to the $\sigma^{+}$ ($\sigma^{-}$) polarization.
The other non-selected transition is barely visible in both cases.
We estimate the purity of polarization to be 99\%,
by calculating $(d_{+} - d_{-})/(d_{+} + d_{-})$, where $d_{\pm}$ is the ODMR depths of the respective $m_S$ = 0 $\leftrightarrow$ $\pm$1 transitions.
It is also evident that we are able to control the microwave polarization continuously and arbitrarily by sweeping $\Delta \phi$, crossing the linear polarization at $\Delta \phi$ = 0$^{\circ}$.
This means that the phase shifter in our setup works as a ``microwave polarizer'' in much the same way as an optical counterpart.

\begin{figure}
\begin{center}
\includegraphics{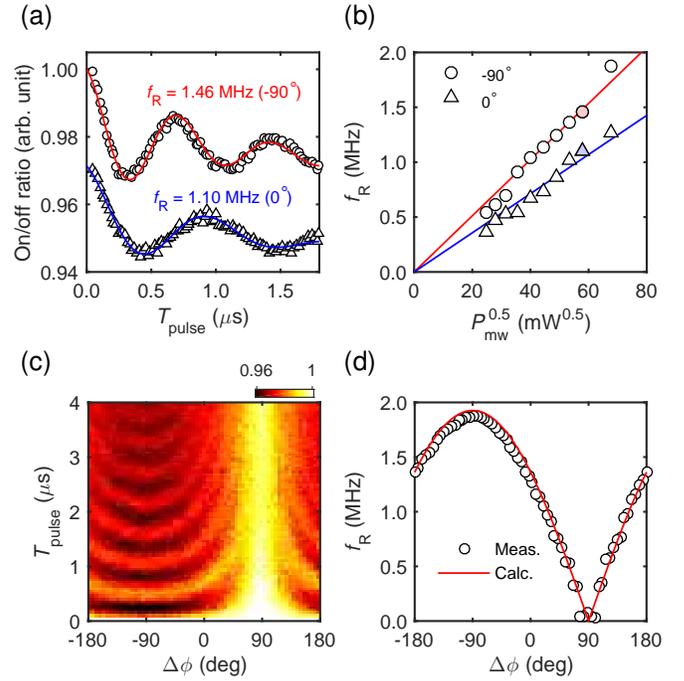}
\caption{(a) Rabi oscillations at $\Delta \phi$ = 0$^{\circ}$ (bottom, shifted by $-$0.03) and at $\Delta \phi$ = $-$90$^{\circ}$ (top).
The solid lines are fits as explained in the main text.
(b) $P_{\mathrm{mw}}$-dependence of $f_{\mathrm{R}}$.
The filled points correspond to (a).
The solid lines are linear fits.
(c) Rabi oscillations as a function of $\Delta \phi$.
The on/off ratio is color coded.
(d) $f_{\mathrm{R}}$ vs. $\Delta \phi$ extracted from (c).
The solid line is a calculation as explained in the main text.
}
\label{fig6}
\end{center}
\end{figure}

Further analysis on the polarization tunability can be made through pulsed ODMR experiments.~\cite{SMS+16}
For a pulsed operation, we apply a microwave pulse sandwiched by the laser illuminations for the spin initialization and readout.
By varying the pulse length $T_{\mathrm{pulse}}$, we observe the Rabi oscillations.
Below, the microwave frequency $f_{\mathrm{mw}}$ is fixed at 2.816~GHz, corresponding to the $m_S$ = 0 $\leftrightarrow$ $-$1 transition. 
Exemplary Rabi oscillations at $\Delta \phi$ = 0$^{\circ}$ and $-$90$^{\circ}$, under the same microwave power ($P_{\mathrm{mw}}$ = 3.33~W), are shown in Fig.~\ref{fig6}(a).
$P_{\mathrm{mw}}$ is measured right before Port~1, and it is also confirmed that nearly the same power is supplied to Port~2.
The Rabi frequency $f_{\mathrm{R}}$ is deduced by either direct fit to or fast Fourier transformation (FFT) of the experimental data.
The curves in Fig.~\ref{fig6}(a) are the fits by
$a \exp(-T_{\mathrm{pulse}}/T_{\mathrm{d}}) \cos (2 \pi f_{\mathrm{R}} T_{\mathrm{pulse}} + \varphi) + b \, T_{\mathrm{pulse}} + c$,
where $a$, $b$, and $c$ are the constants, $T_{\mathrm{d}}$ characterizes the decay time of the oscillation, and $\varphi$ is the phase.
The linear term $b \, T_{\mathrm{pulse}}$ is empirically introduced to better reproduce the experimental data, but is not essential to determine $f_{\mathrm{R}}$.

We go on to take the Rabi oscillations at various $P_{\mathrm{mw}}$, and the values of $f_{\mathrm{R}}$ deduced from them are plotted in Fig.~\ref{fig6}(b) as a function of $\sqrt{P_{\mathrm{mw}}}$.
The linear relation between $f_{\mathrm{R}}$ and $\sqrt{P_{\mathrm{mw}}}$, generally expected for the Rabi experiments, is verified in both cases.
When $\bm{B}_{\mathrm{ac}}$ and $\bm{n}_{\mathrm{NV}}$ are orthogonal as in the present case,
$B_{\mathrm{ac}}$ and $f_{\mathrm{R}}$ are related by $f_{\mathrm{R}} = \sqrt{2} \, \beta \gamma B_{\mathrm{ac}}$,
where the $\Delta\phi$-dependent factor $\beta$ is given by~\cite{beta}
\begin{equation*}
\beta(\Delta \phi) = \frac{\sqrt{1-\sin (\Delta \phi)}}{2}.
\end{equation*}
The two slopes in Fig.~\ref{fig6}(b) are 0.0256~MHz/$\sqrt{\mathrm{mW}}$ for $\Delta \phi$ = $-$90$^{\circ}$ and 0.0178~MHz/$\sqrt{\mathrm{mW}}$ for $\Delta \phi$ = 0$^{\circ}$, giving the ratio of 1.44.
This agrees well with the theoretical ratio $\beta(-90^{\circ})/\beta(0^{\circ})$ = $\sqrt{2}$ $\approx$ 1.41.

The full $\Delta \phi$-dependence of the Rabi oscillation at $P_{\mathrm{mw}}$ = 4.56~W is shown in Fig.~\ref{fig6}(c), and the values of $f_{\mathrm{R}}$ extracted by FFT are plotted in Fig.~\ref{fig6}(d).
We also carried out the 3D electromagnetic simulation at the 4.56~W input power and at $\Delta \phi$ = 0$^{\circ}$, and deduced $B_{\mathrm{ac}}$ = 0.049~mT at the measurement position (inside of diamond).
This corresponds to $f_R$ of 1.932~MHz at $\Delta \phi$ = $-$90$^{\circ}$ [solid line in Fig.~\ref{fig6}(d)], compared with the experimental value of 1.878~MHz (97\% match).
Due to the high agreement between the simulated and measured data, we conclude that arbitrarily polarized microwaves are generated in our resonator circuit with high accuracy.

\begin{figure}
\begin{center}
\includegraphics{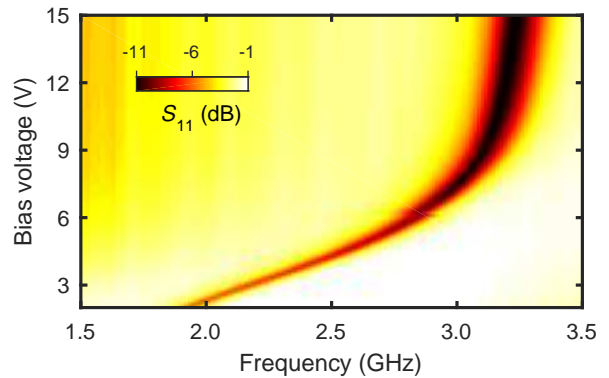}
\caption{Demonstration of frequency tunability.
$S_{11}$ as a function of dc bias voltage applied to varactor diodes (SN: SMV1232) through Port 1.
A bias tee is used to mix the dc and ac signals.
Ports 2 and 4 are equipped with dc blocks.
Ports 3 and 4 are 50-$\Omega$ terminated.
}
\label{fig7}
\end{center}
\end{figure}

Lastly, we demonstrate that not only the microwave polarization but $f_{\mathrm{res}}$ of our resonator circuit is tunable.
For this purpose, we only have to replace the 1.1-pF capacitors with varactor diodes.
As the capacitance of a varactor diode is dependent on an applied reverse voltage, it effectively works as a compact variable capacitor.
As the network analyzer measurements in Fig.~\ref{fig7} demonstrate, higher dc bias voltages on the varactor diodes shift $f_{\mathrm{res}}$ to higher frequencies.
The total frequency range tunable with our varactor diodes is between 2 and 3.2 GHz.
At higher $f_{\mathrm{res}}$, the $Q$ factors decrease slightly, due to larger conductive losses.
Although we demonstrated the frequency tunability of the resonator,
we limit the scope of our ODMR measurement to the fixed capacitors due to their better high frequency performance (ultralow equivalent serial resistance). 

In conclusion, we have presented a microwave circuit design providing spatially uniform, arbitrarily polarized, in-plane oscillating magnetic fields in a wide range of frequencies.
Combined with a (111)-oriented diamond substrate, near-perfect polarization control has been achieved.
In principle, our resonator design is not limited to the NV centers and applicable to other spin-carrying color centers and quantum dots in semiconductors compatible with ODMR.~\cite{TBB+13,W13}
We expect that the design presented in this work thus will be useful not only for advanced study of NV centers in diamond for quantum information and metrology applications
but also for a wide range of magnetic resonance experiments.

KMI acknowledges the support from KAKENHI (S) No.~26220602,
JST Development of Systems and Technologies for Advanced Measurement and Analysis (SENTAN),
JSPS Core-to-Core Program, and
Spintronics Research Network of Japan (Spin-RNJ).
\bibliography{cavity_circular}

\begin{thebibliography}{25}
\bibitem [{bet()}]{beta}%
  \BibitemOpen
  \href@noop {} {}\bibinfo {note} {A magnetic field $\bm{B}_{\mathrm{ac}}(t)$
  oscillating at the angular frequency $\omega$ can be described as
  $\frac{B_{\mathrm{ac}}}{\sqrt{2}} [\cos(\omega t) + i \cos(\omega t - \Delta
  \phi )] = \frac{B_{\mathrm{ac}}}{2\sqrt{2}} [ e^{i \omega t } (1 + i e^{-i
  \Delta \phi}) + e^{-i \omega t} (1 + i e^{i \Delta \phi})]$. Invoking the
  rotation wave approximation, we drop the counter propagating component ($e^{i
  \omega t }$). The absolute value of the $\sigma^-$ ($e^{-i \omega t }$)
  component determines $\beta$.}\BibitemShut {Stop}%
\end{thebibliography}
\end{document}